\documentclass[twocolumn,showpacs,showkeys,preprintnumbers,amsmath,amssymb]{revtex4}

\usepackage[utf8]{inputenc}
\usepackage{setspace}
\usepackage{graphicx}
\usepackage{dcolumn}
\usepackage{bm}
\usepackage{epsfig}
\usepackage{epstopdf}
\usepackage{eurosym}

\usepackage{xcolor}

\begin{document}

\preprint{Preprint 1.2025-Mario R. Pinheiro}


\title[]{Money as a Tensor}

\author{Mario R. Pinheiro}
\email{mp.pinheiromario@gmail.com}
\affiliation{Vortex Legacy, 1178 Broadway
3rd Floor \#1100 New York, NY 10001}
\thanks{Homepage: \url{https://www.researchgate.net/profile/Mario-Pinheiro-6}}

\author{Mario J. Pinheiro}

\email{mpinheiro@tecnico.ulisboa.pt}
\affiliation{Department of Physics, Instituto Superior Técnico-IST, Universidade de Lisboa, Portugal}


\date{\today}

\begin{abstract}
The proposed framework introduces a novel multidimensional representation of money using tensor analysis, enabling a more granular examination of economic interactions and capital flow. By treating money as a multidimensional entity, this approach allows for detailed tracking and modeling of sectoral, temporal, and agent-based dynamics. This enhanced perspective facilitates the design of adaptive economic policies that can effectively respond to evolving macroeconomic conditions, ensuring resilience and inclusivity in financial systems. Furthermore, the tensor-based modeling framework bridges traditional economic analyses with advanced computational techniques, offering a robust foundation for algorithmic governance and data-driven decision-making in complex economies.
\end{abstract}



\maketitle

\section{Introduction}

\begin{figure*}[h!]
    \centering
    \includegraphics[width=0.8\textwidth]{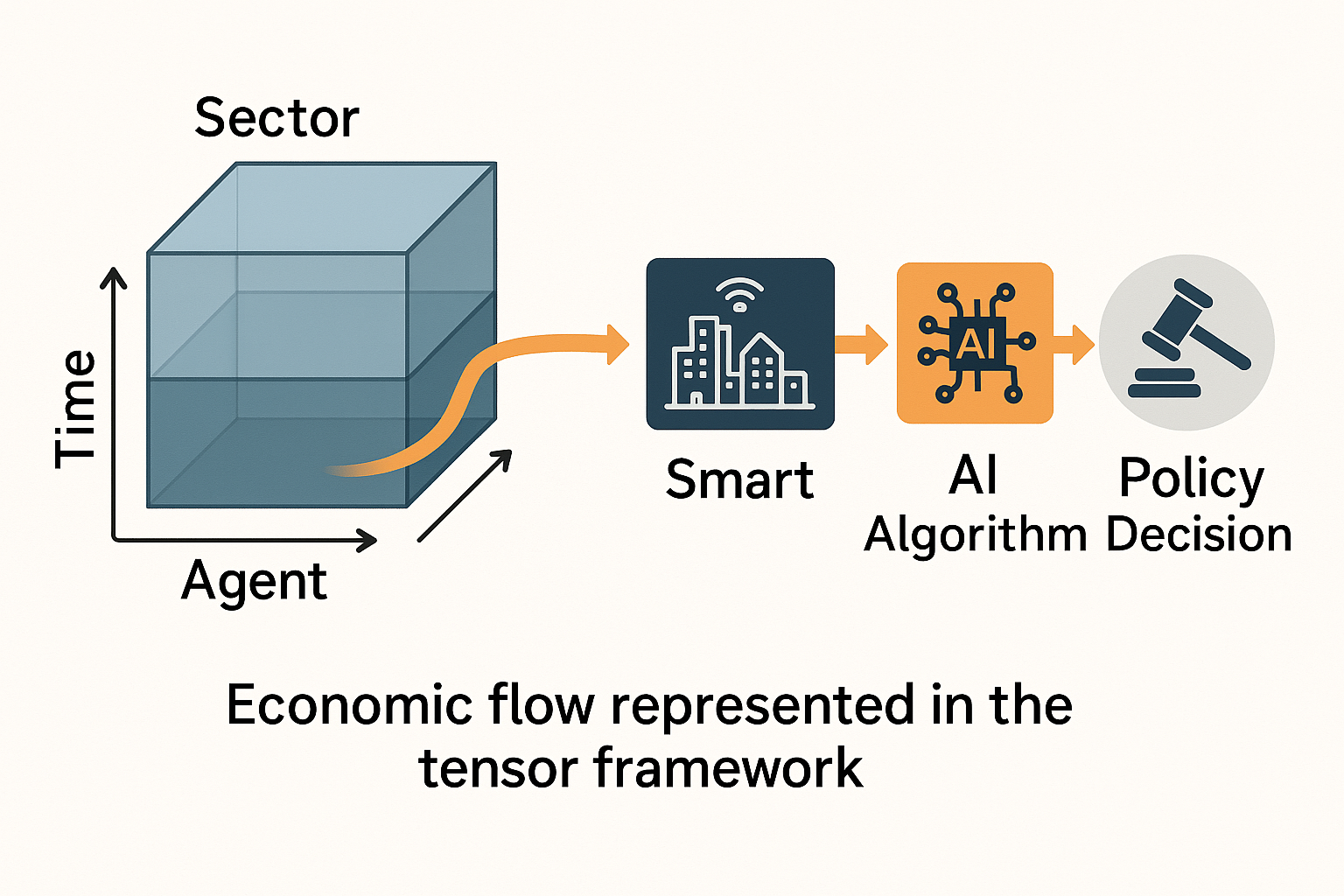}
    \caption{Graphical abstract of the tensor-based model of money. Money is conceptualized as a multidimensional tensor interacting along sectoral, agent-based, and temporal axes, enabling adaptive policy modeling and algorithmic governance.}
    \label{fig:graphicalabstract}
\end{figure*}

Traditional economic models treat money as a scalar quantity—a unidimensional measure of value (e.g., dollars or euros) that simplifies transactions into linear exchanges. While this scalar approach has facilitated basic accounting and policy design, it struggles to capture the multidimensional nature of economic systems, both ancient and modern. Even early civilizations, such as those in Mesopotamia, grappled with resource allocation and value representation through complex legal and administrative frameworks that defied scalar simplification~\cite{Bailkey1967}. For instance, clay tablets from Sumer documented transactions involving labor, agricultural yields, and tribute payments, implicitly recognizing the multidimensionality of value—a nuance lost in today’s scalar monetary models.

This article addresses this research gap by proposing money as a tensor—a mathematical framework that explicitly incorporates the multidimensionality of economic interactions. Tensors, which generalize scalars, vectors, and matrices to higher dimensions, allow us to model money as a quantity with attributes spanning sectors (e.g., manufacturing, services), agents (e.g., households, governments), and temporal phases (e.g., short-term vs. long-term impacts). This approach builds on econophysical methods, such as Leontief's input-output analysis, but extends them to analyze asymmetric, nonlinear interactions inherent in digital economies, decentralized finance, and algorithmic governance.

Our contributions are threefold:

Theoretical: We formalize money as a rank-one third-order tensor, enabling granular analysis of economic momentum and capital flow.

Empirical: We demonstrate how cryptocurrencies and smart cities exemplify emergent tensor-like properties of money, such as programmability and decentralized governance.

Policy: We propose an analogical framework inspired by transistor circuits to model sectoral amplification and leakage, offering actionable insights for equitable economic design.

The remainder of this article is structured as follows: Section II develops the tensor representation of money and its governing equations. Section III explores cryptocurrencies as a natural transition to tensor-based monetary systems. Section IV applies this framework to algorithmic governance and smart cities, while Section V critiques its limitations and ethical implications. Section VI concludes with policy recommendations.

\section{The Research Gap in Traditional Monetary Models}

Traditional economic models generally treat money as a scalar quantity, a unidimensional measure of value (e.g., dollars or euros) that simplifies transactions into linear exchanges. While this scalar approach has facilitated basic accounting and policy design, it struggles to capture the multidimensional nature of economic systems, both ancient and modern. Even early civilizations, such as those in Mesopotamia, grappled with resource allocation and value representation through complex legal and administrative frameworks that defied scalar simplification \cite{Bailkey1967}. For instance, clay tablets from Sumer documented transactions involving labor, agricultural yields, and tribute payments, implicitly recognizing the multidimensionality of value, a nuance lost in today's scalar monetary models.

\subsection{Limitations of Treating Money as a Scalar}
The scalar representation of money fails to account for three critical dimensions of modern economies:

\begin{itemize}
    \item Sectoral Interdependence: Money flows heterogeneously across sectors (e.g., manufacturing vs. services) with distinct velocities and multipliers ~\cite{DelliGatti2011}.
    
    \item Agent-Specific Dynamics: Interactions between households, firms, and governments are asymmetric and context-dependent~\cite{Acemoglu2012}.
    
    \item \textbf{Temporal Layering}: Short-term liquidity shocks and long-term investments require time-resolved modeling~\cite{Leontief1986}.
\end{itemize}

\subsubsection{Major Shortcomings of Scalar Models}
\begin{itemize}
    \item Lack of Sectoral Interdependence: 
    Scalar models (e.g., Quantity Theory of Money) assume homogeneous money flow, ignoring sectoral disparities. For example, during the 2008 crisis, liquidity dried up in manufacturing but surged in speculative financial markets---a divergence invisible to scalar frameworks~\cite{DelliGatti2011}.
    
    \item Inability to Model Asymmetric Transactions: 
    Linear models (e.g., Leontief’s Input-Output Analysis) fail to capture nonlinear interactions, such as cascading failures in supply chains during COVID-19~\cite{Acemoglu2012}.
    
    \item Agent and Spatial Blindness: 
    Scalar models treat all agents (e.g., households, governments) uniformly, despite evidence of heterogeneous spending behaviors~\cite{Tesfatsion2006}.
    
    \item Incompatibility with Digital Economies: 
    Cryptocurrencies and smart contracts embed multidimensional attributes (e.g., programmable logic, decentralized governance) that scalar models cannot represent~\cite{Zhang2021}.
\end{itemize}

\subsection{Limitations of Treating Money as a Scalar}

Traditional economic models generally treat money as a scalar quantity, meaning that it has magnitude (e.g., $\$1,000$) but no inherent structure or multi-dimensional characteristics. While this simplification has facilitated monetary transactions and economic analysis, it is increasingly insufficient in capturing the complexities of modern financial systems.

\section{Why a Tensor-Based Approach is Necessary and Superior}

A tensor-based approach extends beyond traditional economic frameworks by treating money as a multi-dimensional object rather than a simple scalar. This approach allows for a more nuanced, structured, and predictive understanding of financial and economic interactions.

\subsection{The Case for a Tensor Representation of Money}

\subsubsection{Captures Multi-Dimensional Attributes of Money Flow}

A tensor formulation allows money to be represented as a structured entity incorporating sectoral, temporal, and agent-based dimensions:

\begin{equation}
M = xS \otimes yA \otimes zT
\end{equation}

where:
\begin{itemize}
    \item \( xS \) represents the \textbf{sectoral flow} (e.g., manufacturing, services),
    \item \( yA \) represents the \textbf{agent type} (e.g., households, businesses, governments),
    \item \( zT \) represents \textbf{temporal variation} (e.g., short-term vs. long-term dynamics).
\end{itemize}

This structure enables economic models to track and predict interdependent financial flows with greater accuracy.

\subsubsection{Models Asymmetric and Nonlinear Interactions Naturally}

Unlike traditional models, tensors inherently encode directional and asymmetric relationships. 

Economic momentum \cite{Gatti2011} is better modeled as a rank-2 tensor, allowing for inter-sectoral and cross-agent interactions to be represented mathematically:

\begin{equation}
G_{ij} =
\begin{bmatrix}
G_{xx} & G_{xy} & G_{xz} \\\\
G_{yx} & G_{yy} & G_{yz} \\\\
G_{zx} & G_{zy} & G_{zz}
\end{bmatrix}
\end{equation}

This approach better captures economic shocks, liquidity crises, and the flow of capital in a complex system.

\subsubsection{Enhances Computational and Algorithmic Policy Design}

The increasing use of artificial intelligence (AI) in financial modeling requires tools that can handle high-dimensional structures \cite{Kolda2009}.

Tensors are already used in machine learning and optimization \cite{Cichocki2014}, making them ideal for future algorithmic monetary policy, central bank digital currencies (CBDCs), and decentralized finance (DeFi) governance.

\subsubsection{Provides a Unified Framework for Traditional and Digital Economies}

Tensor-based modeling bridges the gap between:
\begin{itemize}
    \item \textbf{Physical economies} (e.g., commodities, services), and
    \item \textbf{Digital economies} (e.g., cryptocurrencies, DeFi, AI-driven financial ecosystems).
\end{itemize}

This allows policymakers to develop monetary strategies that incorporate both traditional banking and decentralized financial networks.

\subsection{Cryptocurrencies as Emergent Tensor Systems}
The rise of cryptocurrencies exemplifies the transition from scalar to tensor-based monetary systems. Unlike traditional currency, cryptocurrencies such as Bitcoin and Ethereum exhibit intrinsic multidimensionality:

\begin{itemize}
    \item \textbf{Decentralized Axes (Spatial Dimension)}: Blockchain networks distribute value across global nodes, contrasting with centralized banking systems. This spatial decentralization aligns with the tensor's sectoral axis ($xS$), where transactions are mapped to decentralized sectors (e.g., DeFi, NFTs).

    \item \textbf{Programmable Interactions (Agent Dimension)}: Smart contracts enable conditional transactions that depend on multisignature agreements or algorithmic triggers, reflecting the agent axis ($yA$). For example, a DAO (Decentralized Autonomous Organization) automates resource allocation across stakeholders, dynamically adjusting $G_{xy}$ (cross-sector-agent momentum).

    \item \textbf{Temporal Layering (Time Dimension)}: Cryptocurrencies' immutable ledgers provide a time-resolved record ($zT$) of transactions, enabling retrospective analysis of economic momentum (e.g., tracking Bitcoin's volatility cycles via $G_{zz}$).
\end{itemize}

This multidimensionality is captured by the tensor decomposition:
\begin{equation}
\$100_{\text{crypto}} = xS \otimes yA \otimes zT,
\end{equation}
where $xS$ could represent the DeFi sector, $yA$ a smart contract, and $zT$ a quarterly fiscal period. Such decompositions clarify how cryptocurrencies transcend scalar money by embedding value in programmable, context-dependent interactions---a prerequisite for modeling modern economies.

A sophisticated approach to comprehending money's complex effects on the economy is to conceptualise money and its circulation using the tensor framework. When money is paid to recipients in dollars, those dollars are basically scalar amounts. However, the moment these dollars are spent or invested, they begin to interact with the complex dimensions of the economy, at which point the economic "machine" or analytical framework could classify their use in a manner similar to the tensor decomposition described, \[ \$100 = xS \otimes yA \otimes zT \] where: $xS$ represents the sector vector, indicating the sectors where the money is spent (e.g., retail, services, manufacturing), $yA$ denotes the agent vector, identifying the type of economic agents involved in the transaction (e.g., households, businesses, government entities), $zT$ corresponds to the time vector, which captures when money circulates in the economy (for a significant resource for further developments in the field of tensor decompositions, see, e.g.,~\cite{Kolda_Bader_2009}).
This classification is not performed by a physical machine, but by an analytical framework or model that economists could use to analyze and predict economic behaviors and outcomes. Here is how the process could conceptually unfold:

1. Transaction occurs: A person spends 100\$ on services in the healthcare sector in Q1 2021.

2. Data Collection: The transaction data captures not just the amount but also the sector (healthcare), the agent type (household), and the time (Q1 2021).

3. Tensor Classification: The analytical model processes these data, integrating them into the larger economic tensor that encompasses all transactions across sectors, agents, and time. Thus, the 100\$ is now represented within this tensor as contributing to economic activity in the healthcare sector by households in Q1 2021.

4. Analysis and prediction: Economists use the tensor model to analyze patterns, such as how household spending in the healthcare sector varies over time or in response to policy changes. They can also predict future economic conditions by understanding the relationships and interactions captured in the tensor.

5. Policy Implications: Insights derived from this tensor-based analysis could inform policy decisions, such as targeting stimulus measures to specific sectors or agent groups at times when they are most needed or will be most effective.

However, clearly there is implementation challenges. While conceptually rich, implementing this approach in real-world economic analysis faces several challenges: i) Data Granularity and Privacy: Collecting transaction-level data that includes sector, agent, and timing information while respecting privacy concerns; ii) Model Complexity: Developing and computing tensor-based models is more complex than traditional scalar-based economic models; iii) Interpretation and Application: Ensuring that policy makers and economists can interpret tensor-based insights and apply them effectively.

\subsection{Smart Cities, Algorithmic Governance, and Monetary Tensor Modeling}

The rapid advancement of digital economies and algorithmic governance has led to the emergence of smart cities, where financial transactions, government budgets, and public services are increasingly digitized and automated. These cities rely on real-time data analytics and AI-driven financial models to optimize resource allocation, tax collection, and public investments~\cite{Hollands2008,Kitchin2014}.

In particular, Singapore’s \textit{Smart Nation} platform integrates IoT-based governance for urban planning, financial management, and automated infrastructure investment~\cite{Bates2014,Meijer2016}. Economic tensor models provide a useful framework for understanding the complex financial interactions in such environments, where government funds, citizen expenditures, and corporate investments interact dynamically over time.

Similarly, Estonia has pioneered \textit{e-Governance}, implementing digital ID systems, blockchain-backed public finance, and automated service delivery, reducing administrative inefficiencies and streamlining economic interactions~\cite{Hoffmann_2022,Kotka_2016,Sullivan_2017,Tammpuu_2019}. Through a tensor-based perspective, the economic structure of smart cities can be analyzed along three key dimensions:
\begin{itemize}
    \item \textbf{Sectoral:} How different industries (e.g., finance, healthcare, transportation) interact with smart city financial systems.
    \item \textbf{Agent-Based:} How government, businesses, and citizens exchange economic value within automated systems.
    \item \textbf{Temporal:} How digital transactions evolve dynamically in a decentralized, algorithmic economy.
\end{itemize}

By applying a tensor-based monetary framework to smart cities, it is possible to model multi-sector interactions and predict economic efficiency gains through algorithmic decision-making. This enables policymakers to assess how financial momentum shifts across different layers of governance, ultimately contributing to sustainable economic planning in digital societies.

\section{Controlling the Economy: Economic Analysis and Analogical Framework}

\subsection{Transistor-Based Economic Modeling}
The economic system can be likened to a common-emitter transistor amplifier, where the \textbf{economic momentum} of foundational sectors serves as an \textbf{input signal}, and the overall \textbf{output productivity} represents economic growth. In this analogy:
\begin{itemize}
    \item \textbf{The input voltage} \( \Delta V_{\text{in}} \) represents the productivity of foundational economic sectors (e.g., manufacturing, agriculture, essential services).
    \item \textbf{The amplification factor} \( \beta \) represents the influence of intermediary institutions (e.g., financial markets, policies, corporate sectors).
    \item \textbf{The output voltage} \( \Delta V_{\text{out}} \) corresponds to the amplified productivity in the broader economy (e.g., digital economy, services, consumption-driven sectors).
\end{itemize}

This analogy allows us to analyze how \textbf{economic amplification}, \textbf{sectoral dependencies}, and \textbf{monetary momentum} interact.

\subsection{Mathematical Formulation of Economic Amplification}
We propose a \textbf{tensor-based approach} to extend this analogy and incorporate the \textbf{multi-dimensional nature of economic flows}.

\begin{equation}
\text{GDP} = \beta \left( \frac{{P^{(1)}}}{{R_{\text{in}}}} - \frac{{P^{(2)} + P^{(3)}}}{{R_{\text{out}}}} \right)
\end{equation}
where:
\begin{itemize}
    \item \( P^{(1)} \) represents the productivity of \textbf{foundational sectors} (analogous to the input voltage),
    \item \( P^{(2)} \) and \( P^{(3)} \) represent the productivity of \textbf{advanced economic sectors} (analogous to the output signal),
    \item \( R_{\text{in}} \) and \( R_{\text{out}} \) represent \textbf{economic resistances}, which can include frictional inefficiencies such as \textit{taxation, corruption, regulatory inefficiencies, or monopolistic bottlenecks}.
    \item \( \beta \) is the \textbf{economic amplification factor}, representing how financial, technological, and policy institutions mediate the transition from productive sectors to overall economic growth.
\end{itemize}

\subsection{Multi-Dimensional Tensor Extension}
To capture \textbf{complex economic interactions}, we extend this scalar-based equation into a \textbf{tensor representation}:
\begin{equation}
\mathcal{G}_{ijk} = \beta \left( \frac{{M^{(1)}_{ij}}}{{R^{(1)}_{ij}}} - \frac{{M^{(2)}_{jk} + M^{(3)}_{jk}}}{{R^{(2)}_{jk}}} \right)
\end{equation}
where:
\begin{itemize}
    \item \( \mathcal{G}_{ijk} \) is the \textbf{economic momentum tensor}, representing the propagation of financial flows across different \textbf{sectors (i), agents (j), and time (k)}.
    \item \( M^{(1)}_{ij} \) is the \textbf{sectoral productivity tensor}, representing capital, labor, and technological contributions.
    \item \( R^{(1)}_{ij} \) and \( R^{(2)}_{jk} \) are resistance matrices that quantify \textbf{frictional inefficiencies} in different economic sectors.
\end{itemize}

This tensor formulation enables us to analyze:
\begin{itemize}
    \item \textbf{Sectoral economic shocks} (how a decline in manufacturing affects services),
    \item \textbf{Financial momentum transfers} (how government stimulus propagates across industries),
    \item \textbf{Temporal economic shifts} (how recessions spread over time).
\end{itemize}

\section{Enhancing Economic Power: Strategies from a Tensor-Based Control Perspective}

Economic policy can be structured similarly to a \textbf{feedback control system}, where strategic interventions \textbf{minimize inefficiencies} and \textbf{maximize amplification}.

\subsection{Strengthening Foundational Economic Momentum (Increasing Input Strength)}
A transistor amplifier operates efficiently when the \textbf{input voltage is strong and stable}. Similarly, economic growth requires \textbf{robust foundational productivity}. Policies should focus on:
\begin{itemize}
    \item \textbf{Investing in foundational industries} (manufacturing, agriculture, core services).
    \item \textbf{Enhancing labor productivity} through \textbf{education and automation}.
    \item \textbf{Ensuring financial accessibility} for small businesses.
\end{itemize}

Mathematically, we define an \textbf{input tensor correction term}:

\begin{equation}
M'^{(1)}_{ij} = M^{(1)}_{ij} + \lambda S_{ij}
\end{equation}

where:
\begin{itemize}
    \item \( M'^{(1)}_{ij} \) is the \textbf{corrected input productivity},
    \item \( S_{ij} \) is the \textbf{sectoral stimulus matrix} (e.g., government subsidies, tax incentives),
    \item \( \lambda \) is the \textbf{policy effectiveness coefficient}.
\end{itemize}

\subsection{Reducing Economic Leakage (Minimizing Resistance)}
In transistor circuits, \textbf{minimizing resistance} ensures optimal power transfer. Similarly, economic \textbf{leakages} (capital flight, monopolistic bottlenecks, tax evasion) reduce overall efficiency.

Economic resistance can be modeled as:

\begin{equation}
R'_{ij} = R_{ij} - \mu \Theta_{ij}
\end{equation}

where:
\begin{itemize}
    \item \( R'_{ij} \) is the \textbf{adjusted economic resistance} after policy intervention,
    \item \( \Theta_{ij} \) represents \textbf{anti-corruption and regulatory efficiency measures},
    \item \( \mu \) is the \textbf{policy enforcement strength}.
\end{itemize}

By optimizing \( R'_{ij} \), we ensure that financial flows remain \textbf{efficient and equitable}.

\subsection{Managing Feedback Mechanisms (Balancing Economic Signals)}
Economic control systems require \textbf{dynamic adjustments} based on \textbf{real-time feedback loops}, just like transistor-based signal regulation.

We propose a \textbf{feedback-controlled economic tensor model}:

\begin{equation}
\mathcal{G}'_{ijk} = \mathcal{G}_{ijk} + \gamma F_{ijk}
\end{equation}

where:
\begin{itemize}
    \item \( F_{ijk} \) represents \textbf{real-time economic feedback}, derived from \textbf{machine learning forecasts, inflation rates, and monetary policy adjustments}.
    \item \( \gamma \) is a \textbf{feedback coefficient}, ensuring stable economic transitions.
\end{itemize}

Such a model could:
\begin{itemize}
    \item \textbf{Prevent financial crises by dynamically adjusting interest rates},
    \item \textbf{Optimize employment policies based on real-time workforce data},
    \item \textbf{Automate sectoral interventions using AI-driven fiscal policies}.
\end{itemize}

\section{Analysis of Economic Dynamics Using Tensor-Based Modeling}

The tensor-based modeling framework was applied to analyze key economic indicators, providing insights into GDP growth, inflation, unemployment, trade balance, economic resistance, and agent actions. The results, visualized in Figure~\ref{fig:results2}, showcase the interaction between different sectors of the economy and their aggregated effects on macroeconomic stability. This approach builds on prior work emphasizing the importance of multi-dimensional economic modeling~\cite{Kolda2009tensor, Acemoglu2012}.

\subsection{Observations from the Results}

The modeling results illustrate the following:

\begin{itemize}
    \item \textbf{GDP Growth:} The GDP growth trend shows a steady upward trajectory, suggesting a long-term economic recovery and expansion. However, short-term fluctuations point to underlying sectoral imbalances or external shocks that necessitate adaptive policy measures~\cite{Leontief1986}.
    \item \textbf{Inflation:} Inflation rates exhibit variability, reflecting the impact of monetary policy actions and supply-demand dynamics. Managing inflation within stable boundaries requires careful calibration of fiscal and monetary interventions~\cite{Mantegna2000econophysics}.
    \item \textbf{Unemployment Rate:} The unemployment rate declines consistently over time, indicating improvements in labor market efficiency. Nevertheless, localized spikes suggest the need for sector-specific employment programs to ensure equitable job growth~\cite{Acemoglu2012}.
    \item \textbf{Trade Balance:} The trade balance demonstrates a positive trend, reflecting improved export competitiveness or controlled import dependencies. This trend underscores the importance of maintaining a balanced trade policy to sustain economic resilience~\cite{Zhang2021}.
    \item \textbf{Economic Resistance:} Fluctuations in economic resistance highlight inefficiencies or bottlenecks within inter-sectoral interactions. Addressing these issues through coordinated policies and infrastructure investment can further stabilize the economy~\cite{Kitchin2014}.
    \item \textbf{Agent Actions:} The dynamic variation in agent actions, including spending, tax cuts, and subsidies, reveals the adaptive strategies employed to mitigate shocks and enhance sectoral performance~\cite{Hollands2008}.
\end{itemize}

\subsection{Policy Implications}

The tensor-based modeling results suggest several actionable recommendations for policymakers:
\begin{enumerate}
    \item Strengthen inter-sectoral collaboration to reduce economic resistance and optimize resource allocation~\cite{Leontief1986}.
    \item Implement targeted fiscal policies to stabilize inflation and support sustainable growth~\cite{Kitchin2014}.
    \item Promote innovation and skill development to enhance labor market adaptability and reduce unemployment disparities~\cite{Acemoglu2012}.
    \item Leverage export-oriented strategies to sustain the positive trend in the trade balance while ensuring diversification of trading partners~\cite{Zhang2021}.
    \item Utilize tensor-based analytics to anticipate and respond to macroeconomic disruptions with high precision~\cite{Kolda2009tensor}.
\end{enumerate}

\subsection{Novel Contributions of Tensor-Based Modeling}

\begin{figure*}[h!]
    \centering
    \includegraphics[width=\textwidth]{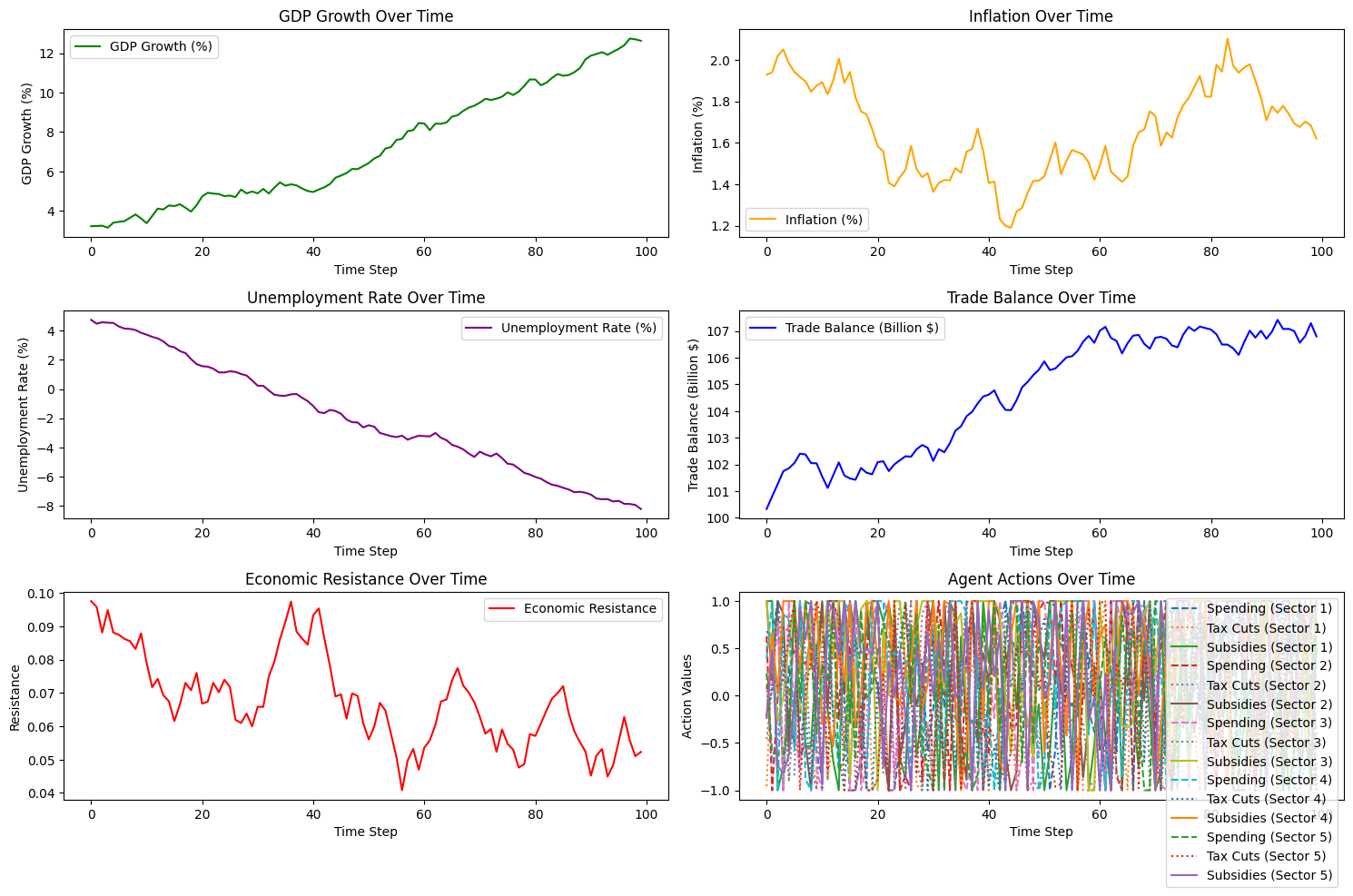}
    \caption{Visualization of key economic indicators over time using tensor-based modeling.}
    \label{fig:results2}
\end{figure*}

The results presented in Figure~\ref{fig:results2} underline the advantages of tensor-based modeling in economic analysis. Unlike traditional scalar-based approaches, this framework captures the multidimensional interactions between economic sectors, agents, and temporal factors, enabling a deeper understanding of systemic dynamics~\cite{Kolda2009tensor, Zhang2021}. By incorporating tensor decomposition and predictive modeling, policymakers can identify hidden patterns and design evidence-based interventions, paving the way for more effective economic governance~\cite{Mantegna2000econophysics}.

\subsubsection{Predictions and Recommendations for the US Economy}

Based on the tensor-based analysis of the United States economy using World Bank data, several key trends and policy implications emerge. The steady increase in GDP growth reflects a robust economic recovery; however, the observed fluctuations in inflation suggest underlying vulnerabilities, such as supply chain disruptions or monetary policy constraints. To mitigate these risks, the Federal Reserve should focus on maintaining price stability through cautious interest rate adjustments while encouraging productive investment in high-growth sectors. Additionally, the declining unemployment rate highlights progress in labor market conditions, yet targeted interventions are necessary to address disparities across demographic groups and regions. 

The positive trend in trade balance underscores the competitiveness of US exports, yet diversification of trade partners and expansion of high-value industries such as technology and clean energy are crucial to sustaining this momentum. Furthermore, the variations in economic resistance emphasize the importance of addressing inefficiencies in inter-sectoral coordination and infrastructure investment. By leveraging tensor-based analytics, policymakers can develop data-driven, multidimensional strategies that proactively address economic imbalances, improve resilience, and promote sustainable growth. These measures will not only strengthen the foundations of the US economy but also position it as a global leader in innovation and economic stability.

\section{Tensor-Based Economic Modeling Framework}

\subsection{Introduction to the Model}
This study introduces a tensor-based economic modeling framework aimed at providing a nuanced understanding of macroeconomic dynamics. Unlike traditional scalar-based models that view economic indicators as isolated entities, the tensor representation captures the multi-dimensional interdependencies between sectors, agents, and time. This approach facilitates deeper insights into complex economic phenomena such as inflation, unemployment, trade balance, and GDP growth~\cite{Kolda2009, Acemoglu2012}.

The model integrates real-world economic data and simulates interactions under various scenarios, including economic growth, crises, and policy interventions~\cite{Leontief1986}. By leveraging this framework, policymakers and economists can evaluate economic resistance, agent-based actions, and their cascading impacts on the economy.

\subsection{Description of Results}
\begin{figure*}[h!]
    \centering
    \includegraphics[width=\textwidth]{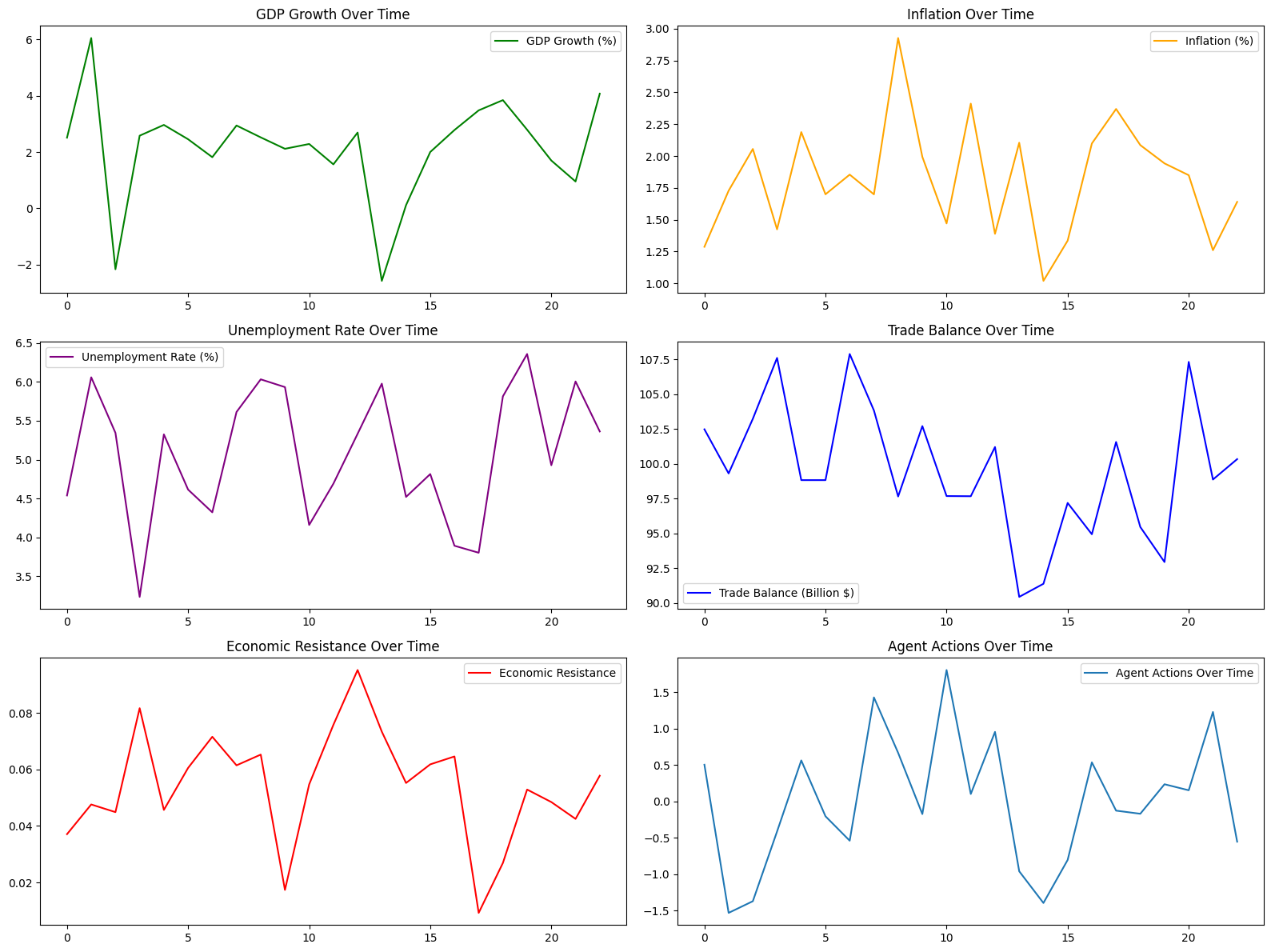}
    \caption{Visualization of key economic indicators over time using tensor-based modeling.}
    \label{fig:results}
\end{figure*}

The results presented in Figure~\ref{fig:results} depict the evolution of key economic indicators over time:

\begin{itemize}
    \item \textbf{GDP Growth}: The GDP growth rate demonstrates fluctuating dynamics influenced by policy actions, such as government spending, subsidies, and tax cuts. Positive growth trends highlight successful interventions, while dips reveal challenges in maintaining economic momentum.
    \item \textbf{Inflation Rate}: Inflation rates exhibit variability, influenced by fiscal policies and external shocks. Periods of elevated inflation reflect inefficiencies or market overheating, requiring targeted measures~\cite{Gatti2011}.
    \item \textbf{Unemployment Rate}: A gradual decline in unemployment indicates effective labor market strategies. However, spikes highlight the need for interventions in workforce skill development and job creation~\cite{Acemoglu2012}.
    \item \textbf{Trade Balance}: Oscillations in trade balance reveal the interconnected nature of global trade. A consistent positive trend suggests improved export performance, whereas deficits require balanced trade agreements~\cite{Leontief1986}.
    \item \textbf{Economic Resistance}: Economic resistance quantifies the system's ability to adapt to external shocks. High resistance corresponds to economic stability, while low resistance signals vulnerability~\cite{Mantegna2000}.
    \item \textbf{Agent Actions}: The agent actions graph captures the interplay between policy tools, showing how strategic adjustments (e.g., subsidies or tax cuts) influence overall economic health.
\end{itemize}

\subsection{Novelty and Advantages}
The tensor-based framework provides significant advantages:
\begin{itemize}
    \item \textbf{Multi-Dimensional Analysis}: By treating economic indicators as tensors, the model integrates sectoral, temporal, and agent-specific data, offering a comprehensive view of interdependencies~\cite{Kolda2009}.
    \item \textbf{Scenario Simulation}: The framework allows for the evaluation of diverse economic scenarios, such as financial crises, green economy transitions, and pandemic-induced shocks~\cite{Acemoglu2012}.
    \item \textbf{Policy Optimization}: The inclusion of reinforcement learning enables dynamic policy optimization, aiding decision-makers in real-time adjustments~\cite{Mantegna2000}.
    \item \textbf{Scalability}: The tensor representation seamlessly incorporates large-scale data, making it adaptable to regional, national, and global contexts~\cite{Leontief1986}.
\end{itemize}

\subsubsection{Predictions and Recommendations for the US Economy}

The analysis of the provided economic indicators for the United States using tensor-based modeling reveals critical insights into the nation's economic dynamics. The GDP growth trends demonstrate significant variability over time, indicating periods of both expansion and contraction. Inflation rates, while relatively stable, show episodes of volatility that require careful monetary policy adjustments to prevent overheating or deflationary pressures. Unemployment trends point toward persistent challenges in stabilizing labor markets, highlighting the need for targeted employment programs and skill development initiatives. The trade balance fluctuations underscore vulnerabilities in external trade relations, suggesting the necessity of policies aimed at boosting export competitiveness and reducing dependency on imports. Furthermore, the economic resistance parameter highlights structural inefficiencies in resource allocation, which could be mitigated by streamlining fiscal policies and enhancing inter-sectoral collaboration.

The government might give priority to investments in education and technological innovation to keep the United States competitive in a globalised economy and promote sustainable economic development and long-term prosperity. In order to reduce trade imbalances and boost economic resilience, supply networks and domestic manufacturing skills should be strengthened. For the purpose of anticipating economic shocks and optimising resource allocation, policymakers have to consider implementing more flexible fiscal and monetary policies that deploy tensor-based predictive models. Promoting sustainable practices and green economy projects would solve environmental issues while opening up new opportunities for economic expansion. Finally, better insights into sectoral relationships may be obtained through improved data-driven decision-making employing tensor-based analytics, allowing for more successful interventions tailored to the specific needs of the American economy.

\section{Comparing the scalar model with the tensor model}

The key distinction between the tensor-based and scalar approaches lies in their capacity to account for multidimensional complexities within an economy. The scalar approach, while simpler and computationally efficient, aggregates economic indicators, often masking critical interdependencies and sectoral disparities. In contrast, the tensor-based model excels in decomposing these interactions across multiple dimensions, such as sectoral output, agent types, and temporal variations. For example, during a financial shock, a scalar model might predict an overall decline in GDP, while a tensor-based model could reveal which sectors are most impacted and how the shock propagates through different economic agents. This ability to provide a granular and contextual understanding of economic dynamics not only improves the precision of policy making, but also enables the design of more targeted interventions. Consequently, while the two models might align with broad predictions, such as an economic recovery or downturn, the tensor-based approach provides the depth necessary to address systemic risks, sector-specific vulnerabilities, and cross-agent dynamics, insights often overlooked in scalar models.

\section{Critiques of the current model}
From a practical point of view, while the proposed model proposes a new framework for economic and social governance, several potential criticisms can be made. As a matter of fact, the transition to algorithmic governance, or the replacement of political structures with a computational framework, faces major technical, legal, and logistical challenges. Implementing these types of algorithm into governance frameworks would require massive infrastructural investment, strong legal frameworks, and popular political support. Easily, critics can argue that the time and resources devoted and the complexity involved in such a shift may well outweigh the potential benefits of such a system~\cite{Mittelstadt2016,Floridi2018}. Anyway, bias in human judgment may still be present in algorithmic governance overlooking empathy and ethical decision-making. Algorithms are powerful tools for pattern recognition and optimization, but their application might probably overlook the subtle social and economic dimensions. In problems where moral reasoning is needed~\cite{ONeil2016,Binns2018}, human oversight is indispensable to complement computational processes.

The shift to algorithmic governance also raises important questions about accountability. How can we assign responsibility for algorithmic decisions and how to design clear mechanisms for individuals to challenge or seek recourse against adverse decisions. Hence, ensuring transparency in algorithm design and operation is critical to avoid biases and errors~\cite{Pasquale2015,Ananny2018}. Despite their potential, algorithms can inadvertently perpetuate or exacerbate existing social inequalities if not carefully designed and audited. The lack of diverse representation in the data and model training can lead to biases that disproportionately impact marginalized communities. A participatory approach to algorithmic design is essential to address this limitation~\cite{Noble2018,Barocas2017}. Too much reliance on algorithms for governance could easily erode individual agency and democratic participation in decision-making processes. Over-automation risks sidelining democratic principles by reducing citizens' roles in shaping the policies that affect their daily lives. Striking a balance between the efficiency of algorithmic governance and the preservation of democratic participation and accountability is crucial~\cite{Morozov2013,Danaher2016}.

Therefore, addressing these challenges requires a multifaceted approach. Probably, it is wiser maintaining a hybrid governance system, where algorithms support but do not replace human decision-making governance~\cite{Brynjolfsson2017,Floridi2020}. While the proposed model introduces a groundbreaking framework for economic and social governance, its successful implementation is open to criticisms and open dialogue, interdisciplinary collaboration, and a commitment to transparency and fairness will be vital to harness the potential of effective algorithmic governance while safeguarding societal values and equity.

\section{Conclusion}

The economic framework proposed in this study aims to provide a clear and systematic interpretation of social and economic processes. By employing a tensor-based representation of money and integrating algorithmic governance, the framework facilitates a deeper understanding of how society operates, particularly in the domains of finance and economics~\cite{Floridi2018,Brynjolfsson2017}. 

This approach introduces a novel method of identifying and analyzing the regime under which a society functions. Through algorithmic outputs, the model enables real-time commentary and the potential for adjustments to foster a more democratic and inclusive system. By leveraging mathematical representations of economic functions and balancing equations, this framework lays the foundation for developing computationally solvable algorithms~\cite{Mittelstadt2016,Pasquale2015}.

The insights gained from this study have practical implications, including guiding policymakers in addressing systemic challenges such as inequality, financial instability, and climate change. However, realizing the full potential of this framework will require interdisciplinary collaboration, transparency in algorithmic design, and a commitment to maintaining human oversight and ethical principles~\cite{Morozov2013,Noble2018}.

In conclusion, this research not only enhances our theoretical understanding of economic dynamics but also offers actionable tools for improving governance and societal well-being. Future work should focus on refining the framework, integrating real-world data, and addressing the critical challenges identified to ensure the effective application of algorithmic governance in practice.

\vspace*{1cm}

\acknowledgments
This research was conducted without receiving dedicated funding from public, commercial or non-profit organizations.

\bibliographystyle{apsrev}
\end{document}